\documentclass{aa}
\usepackage{graphicx}
\usepackage{natbib}
\usepackage{color}
\bibpunct{(}{)}{;}{a}{}{,} 
\begin{document}

\def\ms{M$_{\odot}$}
\def\zs{Z$_{\odot}$}
\def\mi{M$_{\rm IN}$}
\def\zi{Z}
\def\xi{Z$_{\rm IN}$}
\def\me{M$_{\rm ENV}$}
\def\nsn{N$_{\rm Ib,c}$/N$_{\rm II}$}
\def\mup{M$_{\rm Ib,c}$}
\def\lb{L$_{\rm B,\odot}$}

\title{On the relative frequencies of  core-collapse
 supernovae sub-types: the role of progenitor metallicity}
\author{ N. Prantzos \inst{1}          
         \and
         S. Boissier \inst{2}
        }

\authorrunning{Prantzos \& Boissier}
 
\titlerunning{On the relative numbers of supernovae subtypes:
the role of progenitor metallicity}

\offprints{N. Prantzos}

\institute{ Institut d'Astrophysique de Paris, 98bis Bd. Arago, 
              75104 Paris, France,  
              \email{prantzos@iap.fr}
             \and
	     Carnegie Observatories, 813 Santa Barbara Str.,
             Pasadena, Ca 91101, USA, 
              \email{boissier@ociw.edu}
           }
\date{submitted : 2003}

\abstract{We show that the observed ratio \nsn \ of the subtypes
Ib,c and II core-collapse supernovae depends on the metallicity
of the host galaxy, as expected on theoretical grounds. However,
the observed relation differs considerably from expectations
based on non-rotating models of single stars with mass loss.
We argue that the predictions of recent models with rotation offer
a much better agreement with observations, at least for progenitor stars
with solar metallicity; calculations of models with higher and lower
metallicities are required in order to substantiate these conclusions. We also
suggest that systematic surveys of core collapse
supernovae up to redshift $z\sim$1 with the SNAP satellite
would allow to probe the  effect of metallicity on supernovae properties
during the past history of the universe.
}


\maketitle
%

\section{Introduction}

It has been known for sometime now that supernovae with different spectral
signatures and lightcurves, such as IIP, IIL, IIdw, Ib, Ic etc., all explode
by the same mechanism: the gravitational energy released by the collapse
of the Fe core of a massive star is partially
transferred to the stellar envelope and, under certain conditions, 
manages to expell it to space. Despite more than fourty years of intense 
theoretical investigations, the conditions for a succesfull explosion
remain unclear yet (e.g. Janka et al. 2002 for a recent overview).

However, the appearance of the supernova to an external observer, i.e.
its optical spectra and lightcurve, depend only little on the mechanism
of the explosion.
The main factor affecting  the observed features of  core collapse
supernovae is the mass (and the size) of the H-rich envelope of the progenitor
star (see e.g. Hamuy 2003 for a recent overview). 
This depends on the amount of mass loss suffered during the 
star's life. This is, in its turn,  mainly a function of the stellar 
mass and metallicity. [ {\it 
Note:} in the case of binary stars, the interaction of the progenitor with its 
companion may play an even more important role in determining the mass loss.]
It is expected then that the relative numbers of the various core collapse
supernovae subtypes should depend on the metallicity of the host galaxy
(e.g. Maeder 1992, Mowlavi et al. 1998).

In this work we use the well-known metallicity - luminosity relationship for 
late type galaxies (e.g. Garnett 2002 and references therein) to show
that the observed ratio \nsn \ between
supernova of type Ib,c and II does indeed depend strongly on metallicity
(Sect. 3). To our knoweledge, it is the first time that such a relation is
shown to hold observationally.
However, we find that the slope of that relation, as well as the value
of the ratio at solar metallicity, are higher than those expected 
from non-rotating models of single stars with mass loss. 
We show that predictions of recently calculated stellar models with
rotation and mass loss (Meynet \&  Maeder 2003)
offer a quantitative agreement with observations, at least for progenitor
stars with solar initial metallicity. Finally, in Sect. 4 we argue that
systematic surveys of supernovae at redshifts up to $z\sim$1, feasible
with the SNAP satellite, would allow to probe the role of metallicity
on supernova properties during earlier epochs in the history of the Universe.

\section{Types of core collapse supernovae}

In a recent comprehensive study, Heger et al. (2003a) 
suggested the following scheme for the progenitors of 
core collapse supernova types, on the basis of models of single non-rotating 
stars with mass loss (Fig. 1a):





1) For progenitor metallicities \zi$\sim$\zs, 
stars with initial mass \mi$<$\mup$\sim$34 \ms \ keep enough of their
envelope at the end of their lives to explode as SNII (either IIP or IIL).
More massive stars lose completely their H-rich envelope, become WR stars
and explode as SNIb,c.

2) For \zi$>$\zs \ \mup \ decreases, down to $\sim$30 \ms \ at \zi$\sim$3 \zs
(since higher metallicities favour higher mass loss rates through radiation
pressure on the envelope).

3) For \zs$>$\zi$>$0.1 \zs, 
weaker mass loss rates increase \mup, which reaches $\sim$40
\ms \ at  \zi$\sim$0.1 \zs.

4) For \zi$<$0.1 \zs, mass losses are 
negligible and the upper mass limit for SNII is kept at 40 \ms \ (Fig. 1).
That upper limit no more coincides with \mup, since stars in the mass range
\mup$>$\mi$>$40 \ms \ and in that metallicity range  fail to
explode (at least under ``standard'' assumptions of current models, 
i.e. spherical symmetry) and collapse directly to form black holes\footnote{
According to Heger et al. (2003a) non-''normal'' supernovae, powered by jets,
may occur in that mass-metallicity region; but their optical signature
differs from both SNII or SNIb,c.}. Only the most massive stars (see {\it
dotted curve} in upper panel of Fig. 1)
produce a weak SNIb,c explosion in that metallicity range.

A direct test of that scheme was recently undertaken by Smartt et al. (2003).
By using high quality images of nearby host galaxies of SN 
prior to the explosion,
and by comparing to stellar evolution tracks with mass loss but no rotation,
they were able to evaluate (or put upper limits on) the mass of the
progenitor stars of eight core collapse supernovae. Seven of them were
SNII (either P or L) and 
were found  to have \mi$<$25 \ms, in agreement with theoretical expectations.
The last one (SN2002ap), classified as Ic,  was found to have
\mi$<$40 \ms \ and \zi$\sim$0.5 \zs, in (marginal) agreement with the scheme
proposed by Heger et al. (2003a).

In this work we propose an indirect test of the ideas of Heger 
et al. (2003a), concerning the
{\it number ratio of SNIb,c to SNII core collapse supernovae as a function of
metallicity}. For metallicities \zi$>$0.1 \zs, where 
the upper mass limit for SNII coincides with \mup \  this ratio is given by
\begin{equation}
{{N_{\rm Ib,c}}\over{N_{\rm II}}} \ = \ {{\int_{M_{\rm Ib,c}}^{M_{\rm SUP}} \ \Phi(M) \ dM}
\over{\int_{M_{\rm INF}}^{M_{\rm Ib,c}} \ \Phi(M) \ dM}}
\end{equation}
where $M_{\rm INF}\sim$9 \ms \ is the lower mass limit for core collapse supernova
(Heger et al. 2003a),
$M_{\rm SUP}\sim$100 \ms \ is the corresponding upper mass limit (on which the 
value of \nsn \ depends very little) and $\Phi(M)$ is the stellar initial 
mass function.  
Eq. (1) is obviously correct for a ``steady state'' situation,
i.e. for a star formation rate $\sim$constant
during at least the last $\sim$25 Myr (the lifetime of a star with
mass $M_{\rm INF}$); it does not apply e.g.
in the case of a starburst more recent than 10 or 20 Myr.

\begin{figure}
\centering
\includegraphics[width=0.5\textwidth,height=10cm]{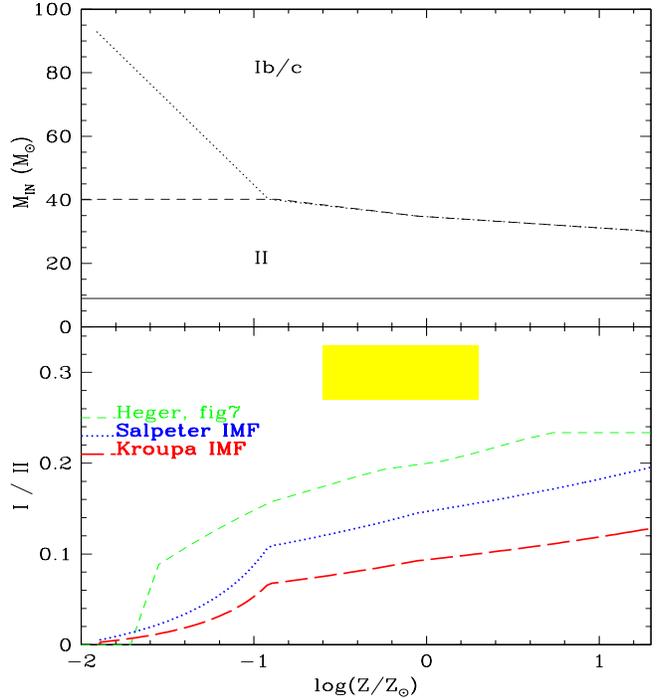}
\caption{\label{f1} 
{\it Top:} Lower mass limit for core collapse supernova ({\it solid curve}),
upper mass limit for  SNII ({\it dashed}) and
lower mass limit for SNIb,c ({\it dotted}) as a function
of metallicity, according to the estimates of Heger et al. (2003a); for 
metallicities higher than 0.1 \zs \ the dotted and dashed curves
coincide, but for lower metallicities stars with
mass 40$<$\mi$<$\mup \ collapse to black holes.
{\it Bottom:} Ratio of \nsn \ as a function of metallicity, taking into account
the mass limits of Fig. 1a; the {\it thick}
curves are obtained with the IMFs of Salpeter ({\it dotted})
and Kroupa et al. (1993, {\it long dashed}), respectively, 
both extended up to 100 \ms. The {\it thin, short-dashed curve} 
is obtained by Heger et al. (2003a) with the 
Salpeter IMF extended to 400 \ms (possibly appropriate for $\sim$zero
metallicity environments). The {\it shaded aerea} corrsponds to the
observed average ratio of \nsn, according to Bressan et al. (2002, lower
value) and Hamuy (2003, upper value), within the metallicity limits
of the host galaxies (see Sect. 3).
} 
\end{figure}

For metallicities \zi$>$0.1 \zs, one sees that
the ratio \nsn \ increases
monotonically with \zi, since the dividing mass \mup \ decreases with
increasing \zi. This is true as far as the stellar initial mass function (IMF)
remains constant with metallicity, which appears to be the case according
to most current studies (e.g. Kroupa 2002). Of course, the value of \nsn \
depends on the form of the adopted IMF in the massive star range, as can 
be seen on Fig. 1b. The two thick curves are constructed, respectively,
with the Salpeter IMF (a single power law with slope X=1.35) and the
Kroupa et al. (1993) IMF (a multi-slope power law with X=1.7 in the massive
star range)
\footnote{The power-law stellar IMF of slope X is defined as 
$dN/dM \ \propto \ M^{-(1+X)}$.}.
In both cases we assume that the IMF extends to $M_{SUP}$=100 \ms.
Indeed, although stars with \mi$>$100 \ms \ might have existed in $\sim$zero 
metallicity environments (Nakamura \& Umemura 2001), 
their existence at non-zero metallicities is
implausible, both on theoretical (e.g. Baraffe et al. 2002) 
and observational grounds (e.g. Heydari-Malayeri 2003).
We assume then in the following that for \zi$>$0.1 \zs \ the IMF extends only
up to $\sim$100 \ms. 

An inspection of Fig. 1b shows that \nsn \ never exceeds $\sim$0.2, even at
the highest metallicities calculated by Heger et al. (2003a) with their
non-rotating models.
This is in disagreement with observations. Indeed, Bressan et al. (2002)
find that in normal galaxies \nsn$\sim$0.27, while Hamuy (2003)
argues that one out of four core collapse supernovae belongs to type Ib,c, 
i.e. that the ratio \nsn \
should be $\sim$0.33.

In the next section we show that,
by counting supernova types (and corresponding ratios) in galaxies of different
metallicities, the prediction of Fig. 1b can be tested statistically.
We find that the predictions of non-rotating models disagree not only
with observed values at high metallicities, but also
with the observed slope of the \nsn \ vs metallicity relation.
We suggest that models including rotation
offer a physically plausible quantitative solution.

\section{The \nsn \ ratio in galaxies}

The existence of a metallicity-luminosity relation is a well established 
fact about spiral galaxies (see Garnett 2002 and references therein).
Oxygen abundances measured in the gaseous phase and at  fixed fractional 
galactocentric radius (at the center, at one disk scalelength or at the 
disk half-radius) always increase with the blue luminosity of the galaxy.
According to Garnett (2002) this relation can be expressed as
\begin{equation}
log \left(\frac{O}{H}\right)=-0.16 {\rm M_B} -6.4
\end{equation}
where M$_{\rm B}$ is the galaxy's blue magnitude and the metallicity is  taken
at the galaxy's effective radius R$_{\rm eff}$. The relation
is valid over $\sim$2 orders of magnitude for O/H and over a range of
$\sim$10 magnitudes, for spirals, irregulars and low surface brightness galaxies
alike (see Fig. 2, top).

This important feature has not found a satisfactory explanation up to now.
It could be due either to a greater efficiency of star formation
in more massive galaxies (e.g. Ferreras and Silk 2001), 
to a younger age for less massive galaxies
(e.g. Boissier et al. 2001), to a more important mass loss reducing the
effective yield in less massive galaxies (e.g. Dekel and Silk 1986) 
or to some combination
of these factors. Independently of its origin, this relation allows
us to test statistically the theoretically predicted trend of \nsn \
with metallicity.

\begin{figure}
\centering
\includegraphics[width=0.5\textwidth]{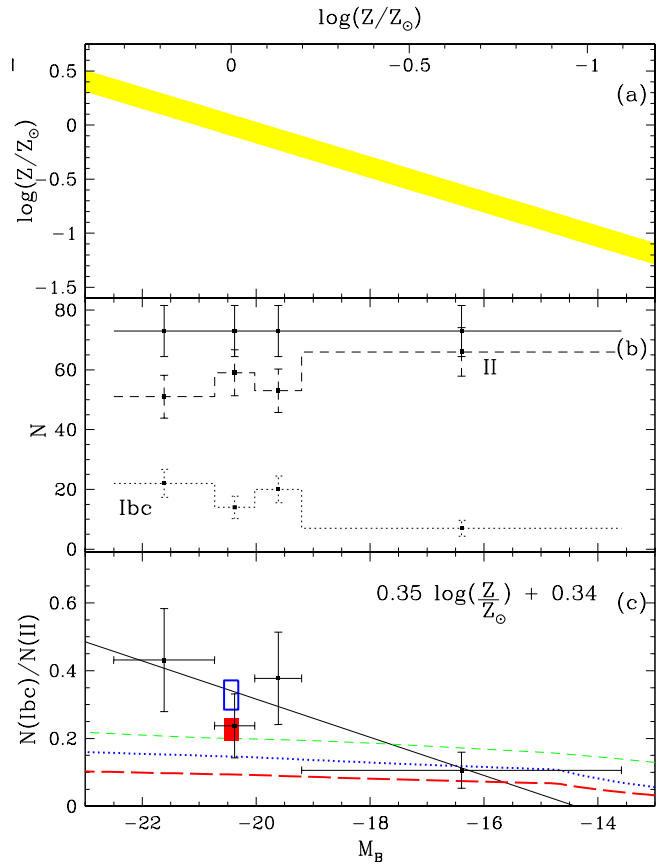}
\caption{\label{f2} 
{\bf Top:} Observed Metallicity-luminosity relation for late type galaxies
according to Garnett (2002) and Eq. (2).
{\bf Middle:} Observed number of core collapse SN sub-types 
as a function of the host 
galaxy's blue magnitude  M$_B$; bins of M$_B$ are such that
the same total number of events (70) are found in each bin; it is found
that the number of SNII ({\it dashed}) decreases, 
while the number of SNIb,c ({\it dotted}) increases
with increasing luminosity. 
{\bf Bottom:} Ratio \nsn \ as a function of blue luminosity of host galaxies;
the observed ratio  (the {\it solid} line is a best fit to the data) 
increases with luminosity, i.e. with metallicity, more rapidly than
theoretical expectations ({\it dotted, long dashed, short dashed} curves,
with same meaning as in Fig. 1b). The {\it filled} and {\it open}
rectanges  at solar metallicity are obtained with the Kroupa et al. (1993)
and the Salpeter IMF, respectively, when \mup \ is decreased
to 22-25 \ms, as appropriate for rotating star models (see text).
}  
\end{figure}

We use a recent version (january 2003) 
of the Asiago Supernova Catalogue,
presented in Barbon et al. (1999)\footnote{Available 
 at {\tt http://merlino.pd.astro.it/\~{}supern/}.}.
This catalogue provides the morphological type 
of the galaxy, the supernovae type and the 
blue magnitude of the parent galaxy when possible
(usually from the RC3 or the LEDA database).
From the entries of the catalogue, we keep only the
galaxies of spiral or irregular morphologies,
(for which the metallicity-luminosity relationship applies) and 
which show no signs of recent starbursts (so that Eq. (1) applies);
we also take into account only the core collapse supernovae, i.e.
those  with a clear identification as one
of the types Ib, Ib/c, Ic, or II. The parent galaxy's 
magnitude is available for 280 of the core-collapse events.
For that restricted sample, we find a value of \nsn=0.27, i.e.
similar to the value given by Bressan et al. (2002).
To study the effect of metallicity on the
number of SN of each sub-type, we adopt magnitude bins
containing each the same number of core collapse events (50).
In the following we adopt errorbars 
$\delta N = \sqrt{N}$ for any number N, and the relative uncertainty of
a ratio $N/M$ is $\delta N / N + \delta M / M$.

In Fig. 2 (middle)  we plot the number of the core collapse
SN sub-types in each bin. 
Despite the poor statistics, it can be seen that the total number of
SNII increases in the  low luminosity bins, while the one of SNIb,c decreases.
This is translated into a clear  increase of the \nsn \ ratio with galactic 
luminosity, presented in Fig. 2 (bottom).

The obvious physical explanation for the observed corelation between 
\nsn \ and M$_B$ involves the effect of metallicity, as explained in
Sect. 2: more luminous and metal-rich galaxies are expected  to have higher
ratio of \nsn. 
Note that the data span the metallicity range 1/3 \zs \ to 2 \zs, 
i.e. the range where theoretical models predict indeed a variation of
\nsn \ with metallicity. 

 However, before discussing further 
the metallicity dependence of \nsn, one should account for the fact that
Eq. (2) is valid at a fixed radius R$_{\rm eff}$, while galaxies display 
important metallicity
gradients and, in consequence, supernova inside a galaxy of a given magnitude 
are expected to occur in regions of different metallicities; this introduces,
in principle, a dispersion of the \nsn \ ratio around some average 
value corresponding to the mean galaxian metallicity. The question is then
whether the observed supernovae exploded in regions with metallicity close to
the one given by  Eq. (2), i.e. close to R$_{\rm eff}$. 
An inspection of Fig. 1 in Bressan et al. (2002)
shows that this is indeed the case:
the distribution of observed core collapse supernovae as a function of their
galactocentric distance peaks at ~0.6 R$_{25}$ (R$_{25}$ is the
radius where the B band surface brightness is 25 mag arcsec$^{-2}$); 
on the other hand, for exponential disks with
``canonical'' central surface brightness ($\mu_0$=21.65 mag/arcsec, the
Freeman value), R$_{\rm eff}$ corresponds to $\sim$0.56 
R$_{25}$\footnote{
For an exponential disk with a scalelength $R_{\rm D}$, the surface brightness
profiles has the form: $\mu(R)=\mu_0+1.085 R/ R_{\rm D}$. 
For $\mu_0$=21.65, $\mu$=25 
occurs at $R=R_{\rm 25} \sim 3 R_{\rm D}$. The radius within which the integral of this
profile is equal to half the integral up to an infinite radius 
is the half-light
radius $R_{\rm eff}$, and is equal to 1.865 $R_{\rm D}$. 
Then, $R_{\rm eff}/R_{25} \sim$ 0.6.}.
Thus, for most of the observed core collapse supernovae,
the magnitude of the host
galaxies can indeed be associated to the metallicities implied by Eq. (2).

A best 
fit to the observed \nsn \ vs metallicity relation is given by
\begin{equation}
{{N_{\rm Ib,c}}\over{N_{\rm II}}} \ = \ 0.35 \ log(Z/Z_{\odot}) \ + \ 0.34
\end{equation}

The theoretical \nsn \ ratio  (i.e the curves
of Fig. 1b) are also plotted in Fig. 2 (bottom), by assuming the
luminosity-metallicity relation of Eq. (2). It can be seen that theory
matches observations only at the bin of lowest luminosity, corresponding
to a metallicity of $\sim$0.3 \zs. At higher metallicities/luminosities there
is an increasing difference between theory and data, by a factor of 2-3 
at \zs  \ and up to 3.5 at $\sim$2 \zs. 
The disagreement never exceeds 2 $\sigma$
but it is worrying that the observed slope is considerably more steep than
the theoretical one.    

Several effects might be at the origin of that discrepancy.
A shallower stellar IMF than the one of Salpeter could certainly
produce a larger \nsn \ ratio; but 
to match the \nsn \ ratio  at \zs, it would
take a slope X$<$1 for the stellar IMF (see Fig. 3), 
i.e. much smaller than all current empirical
evidence. On the other hand, an IMF with a {\it metallicity dependent slope}
could obviously fit to the data. However, 
such a variability of the IMF is not supported
by observations (see e.g. Kroupa 2002 and references therein).

\begin{figure}
\centering
\includegraphics[angle=-90,width=0.5\textwidth]{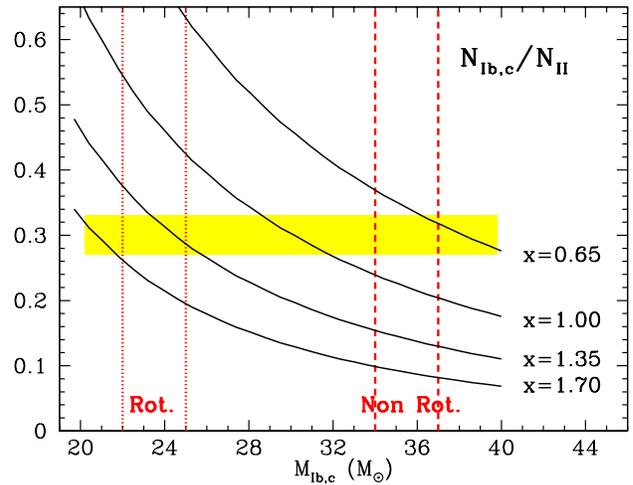}
\caption{\label{f3} 
The  \nsn \ ratio as a function of  \mup \
(the lower mass for a massive star to explode as Ib,c),
for several values of the
slope X of the stellar IMF ($dN/dM\propto M^{-(1+X)}$, so that 1.7 corresponds
to the Scalo IMF and 1.35 to the Salpeter IMF).
The two sets of vertical lines
indicate the values of \mup \ obtained with: 
i) non-rotating models ({\it dashed}),
respectively
at 34 \ms \ from Heger et al. (2003a) and at 37 \ms \ 
from Meynet \&  Maeder (2003);
ii)  rotating models ({\it dotted}) from Meynet \&  Maeder (2003), respectively
at 22 \ms \ (stars become WNE but retain part of 
their H-envelope at their death)
and at 25 \ms \ (stars lose their entire H-envelope). Both cases concern
stellar models of solar initial metallicity.
The {\it  shaded area} indicates the observed range 
of values of  \nsn \ according to Hamuy (2003) and Bressan et al (2002).
}
\end{figure}

As pointed out in Sect. 2, the theoretical predictions are based on the single 
star models of Heger et al. (2003a), calculated with mass loss.
Other ingredients, not taken into account in those models, may affect the
\nsn \ ratio. For instance, mass loss through Roche lobe overflow in a binary
system is a well known channel to form WR stars (e.g. Chiosi and Maeder 1986),
which are progenitors of SNIb,c. This channel was thought 
to be   predominant at
low metallicities: Maeder \& Meynet (1994)
found that all WR stars in the Small Magellanic Cloud
are in binaries, but argued that 
the WR/O ratio in different metallicity environments suggests
that the fraction of the WR stars due to that channel is
small at high metallicities. 
 Recently, Foellmi et al. (2003a) performed a thorough
study of the WR population of the SMC, and found that, contrary to
theoretical expectations, the binary fraction of WR in that galaxy
is $\sim$40$\%$ rather than  $\sim$100$\%$ theoretically
expected. Also  Foellmi et al. (2003b) 
found that the corresponding  value for the LMC 
(which is   $\sim$3  times more metal-rich than the SMC) is  $\sim$30$\%$.
These findings weaken even more the role of binarity in WR formation
at high metallicities. In fact, assuming that the binary channel
contributes by 20$\%$ to the Milky Way population of WR 
(Foellmi et al. 2003a,b),
that binarity reduces the average WR mass to 19 \ms  \ (for solar metallicity
stars) and taking into account the results displayed in Fig. 3, one sees
that the expected \nsn \ value at solar metallicity is 0.23, i.e.
still lower than observed in solar metallicity environments.
Moreover, the
contribution of binaries to the WR fraction and to the \nsn \ ratio should
be independent of metallicity and cannot therefore account for the
mismatch obtained in Fig. 2 (bottom) between observed and theoretical slopes.

Models of {\it rotating massive stars} (combined with mass loss)
have been recently developed by a few groups
(Maeder \& Meynet 2002, Heger et al. 2003b). Meynet \&  Maeder (2003)
find that rotation produces stars with more massive convective cores,
higher effective temperatures and smaller mass left at the end of the
evolution, compared to non-rotating ones. 
In particular, they find that the minimal mass for a star to
become WR is 37 \ms \ for non-rotating stars (i.e. close to the
34 \ms \ suggested by Heger et al. 2003a), but  only 22 \ms \ for stars 
rotating initially at 300 km/s.  In fact, such stars become only WNL stars,
i.e. they keep a trace amount of their 
original hydrogen envelope. An inspection
of Figs. 9 and 10 of Meynet \&  Maeder (2003) shows that 
stars rotating at 300 km/s with slightly higher mass, around 25 \ms,
become WNE stars losing   their entire H-envelope.

This reduction in \mup \  obviously affects the \nsn \ ratio. 
In Fig. 2 (bottom)
 we also plot the \nsn \ ratio assuming \mup=22 and 25 \ms, respectively,
for the Salpeter and for the Kroupa et al. (1993) IMFs and for 
solar metallicity only ({\it open} and {\it filled} squares, respectively).
It can be seen that there is a much better agreement with the
observations. As argued by Meynet \&  Maeder (2003), rotation
provides a better fit to various observations concerning WR stars,
like statistics of WR subtypes and surface abundances. According to
the results displayed in Fig. 2, it also provides a nice 
explanation to the observed \nsn \ ratio, at least for solar
metallicity stars. Indeed, if it is assumed that \mup \ is as high as
34 \ms, it requires unnaturaly low values of 
the slope X of a power-law
IMF in order to obtain the observed \nsn \ ratio, as shown in Fig. 3.
Only for values of \mup \ at least as low as 22-25 \ms, 
the observed \nsn \ ratio can
be reproduced with ``reasonable'' values of X. We suggest
that this is a further argument in favour of
the rotating star models of Meynet \&  Maeder (2003).

Assuming that the fit of Eq. (3) represents well the metallicity dependence
of the \nsn \ ratio, and that the only factor affecting that dependence
is the variation of \mup \ with metallicity, one may use Eq. (1) to determine
that variation. This is done in Fig. 4, for two IMFs. The thick portions
of the curves indicate the region of galaxian metallicity (or blue magnitude)
where \nsn \ has been observationally determined (i.e. between M$_B$ values
of -23 and -14 in Fig. 2c). It can be seen that \mup \ varies from 40-50
\ms \ at Z$\sim$0.3 \zs \ to slightly below 20 \ms \ at Z$\sim$2 \zs.
This observationally determined variation of \mup \ may offer potentially
important constraints to models of massive star evolution.

 We note that Bressan et al. (2002) find that the ratio \nsn \ in Seyfert
galaxies is $\sim$1, i.e. about three to four times larger than in normal
galaxies. They explore several possible reasons for such a high ratio,
including metallicity effects on \mup, and they conclude that
the most probable one is a recent starburst, of age smaller than the
lifetime of the smallest core collapse supernova (in which case Eq. (1)
is invalid). They base their quantitative estimates on non-rotating models
with mass loss. We think their explanation
certainly holds, but rotating models may alleviate the constraints on the
age of the starburst they obtain in their Fig. 3.

\begin{figure}
\centering
\includegraphics[angle=-90,width=0.5\textwidth]{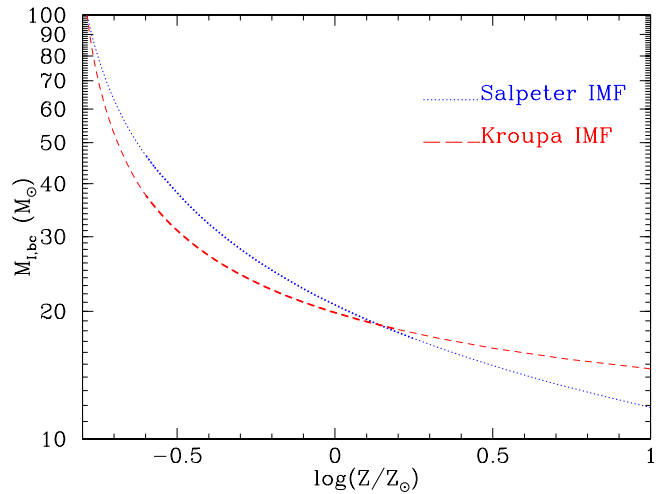}
\caption{\label{f4} 
Mass  \mup \
(the lower mass for a massive star to explode as Ib,c)
as a function of the initial stellar metallicity Z, estimated from Eq. (1)
as to reproduce the values of \nsn=f(Z) corresponding to the fit to the
observational data displayed in Fig. 1c.
}
\end{figure}

\section{Relative supernova frequencies in galaxies}

Supernova frequencies are usually given for galaxies of different morphological
types, in units of SNu (i.e. in number of SN per 100 yr and per 10$^{10}$
\lb). The rational behind this definition of frequency units is that the
galaxy's blue luminosity measures its total mass, as also suggested by
the Tully-Fisher relation in the blue. Note that in starburst galaxies
the blue luminosity traces rather the young stellar
population, while in normal
galaxies there is a significant contribution from the old, metal poor,
population.

As expected then, the SN frequencies expressed in SNu depend little
(if any at all) on the galaxy's morphological type (e.g. Cappellaro
et al. 1999, Turatto 2000), as shown on Fig. 4a. For the same reason,
the ratio \nsn \ is also insensitive to the galaxy's morphological type
(Fig. 4b) and it is of little utility in galaxy studies.
We only plot those ratios in Fig. 4b as a consistency check of our
results with those of previous surveys.
On the other hand, Cappellaro et al. (1999) 
have shown that the frequencies of core collapse supernovae (II+Ib,c)
depend significantly on the galaxy's U-V colour, in the sense that
higher frequencies correspond to smaller U-V values, i.e. to younger
galaxies; they suggested that this dependence reflects directly the
dependence of the core collapse SN rate on the current
star formation rate.
If confirmed, this finding would allow then to probe the star formation
rates of late type galaxies, i.e. their current status
(provided that the relevant uncertainties, that is the effect of extinction
on U-V indices, are adequatly taken into account).

\begin{figure}
\centering
\includegraphics[width=0.5\textwidth]{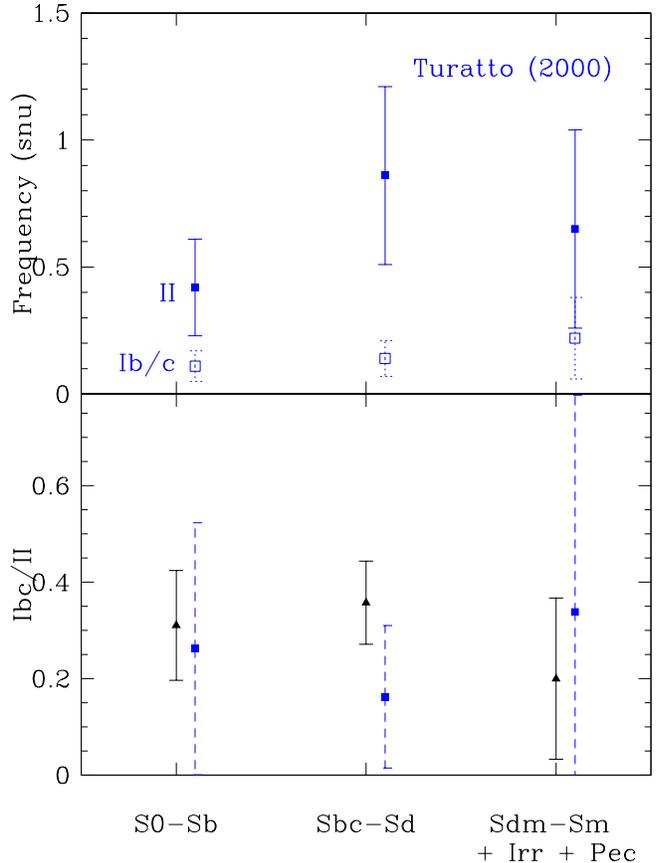}
\caption{\label{f5} 
{\it Top:} Frequency of SNII and SNIb,c (expressed in SNu, i.e.
number of SN per 100 yr and per 10$^{10}$ L$_{B,\odot}$)  for various
galaxy morphological types (from Turatto 2000).
{\it Bottom:} Ratio \nsn \ in galaxies of different morphological types,
according to Turatto (2000, {\it dashed}) 
and from the catalogue used in this work ({\it solid}); the agreement is
fair for S0-Sb and later than Sdm types, and satisfactory for Sbc-Sd
galaxies, although the uncertainties are much higher in the case
of Turatto (2000; this is
due to the fact that he evaluates SN subtype 
{\it frequencies} (and thus requires 
control exposure times, with associated uncertainties), while
we take only subtype {\it ratios}.
} 
\end{figure}

We suggest here that instead of SN frequencies, {\it ratios} between subtypes
of core collapse supernovae coud be used as a probe of the metallicity
of the host galaxies, which is a measure of the integrated star formation
activity of the galaxy, i.e. of its {\it past history}.
Measurements of numbers of SN  subtypes at higher redshifts would then allow to
probe in much greater detail the history of metallicity evolution in the
Universe. For that purpose, it would not be sufficient to simply plot 
the \nsn \ ratio as a function of redshift (as in e.g. Heger et al. 2003a).
A detailed survey,  equivalent to the one presented in Fig. 2,
would be necessary: plotting the \nsn \ ratio as a function of the
(rest frame) blue magnitude of galaxies at 
different redshifts, would enable us
to check whether the variation of that ratio with blue magnitude 
was the same in the past; since this relationship is an indirect one (i.e 
via the metallicity of the host galaxy), such a survey would offer also
clues about the metallicity-luminosity relation in the past.

Such a detailed survey would be possible with future instruments, such as
the SNAP satellite (see web page at: {\tt
http://snap.lbl.gov/}). SNAP is expected to detect $\sim$2000 SNIa
up to redshifts of 1.7.
Taking into account that core collapse supernovae are fainter (by 
$\sim$ two magnitudes) than SNIa and 2-3 times as frequent, one sees that
SNAP will be able to construct large enough samples  
for the purposes of the proposed study at
several redshift ranges up to z$\sim$1.

\section{Summary}
This work explores (some of) the consequences of metallicity dependent 
mass loss of massive stars 
on the statistics of core collapse supernovae of various subtypes,
and in particular on the ratio of type Ib,c vs. type II supernovae.
According to current models of massive stars, 
this ratio should increase with increasing metallicity (Heger et al. 2003a).
The exact form of that relationship depends (a little) on the adopted
stellar IMF and mostly on the value of \mup, the minimal value of initial
mass for a star to explode as SNIb,c.

We show first that the observed \nsn \ does indeed display a metallicity
dependence. For that purpose we construct a statistically significant sample
of Ib,c and II supernovae as a function of their host galaxy's blue
magnitude and we use the well-known metallicity-luminosity relation for 
late type galaxies (Garnett 2002).

However, the observed \nsn \ vs Z relationship is steeper than theoretically
expected, if one adopts non-rotating star models (from e. g. Heger et al. 2003a
or Meynet \& Maeder 203).
We argue that neither the assumptions of 
a metallicity dependent IMF or of an enhanced fraction of Ib,c supernovae originating
in binary systems offer viable solutions to the problem.
We suggest that rotating stellar models with mass loss, such as those
recently calculated by Meynet \&  Maeder (2003) offer a much better
quantitative explanation. We base our argument on models of solar initial
metallicity and we urge calculations at lower {\it and} higher than solar
metallicities in order to substantiate our conclusions.

Furthermore, we suggest that surveys of core collapse supernovae
with the  SNAP satellite will allow to probe the
effect of metallicity on supernovae properties during a large fraction
of the past history of the Universe, at least up to redshift $z\sim$1.

\acknowledgements{ We are grateful to the referee, G. Meynet, for his critical 
comments and suggestions, which improved the paper substantially.}

\def\aj{AJ}
\def\apj{ApJ}
\def\apjs{ApJS}
\def\aap{A\&A}
\def\aaps{A\&AS}


{}

\end{document}